\documentclass[11pt, a4paper, notitlepage]{article}

\usepackage{amsmath, latexsym, amsthm, amsfonts, amssymb} 
\usepackage{graphicx} 
\usepackage{epsfig}
\usepackage{varioref}

\theoremstyle{plain}
\newtheorem{thm}{Theorem}[section]
\newtheorem{prop}{Proposition}[section]

\newtheorem{lem}{Lemma}[section]

\theoremstyle{definition} 

\makeindex

\newcommand{\beao}{\begin{eqnarray*}}
\newcommand{\eeao}{\end{eqnarray*}\noindent}

\newcommand{\beam}{\begin{eqnarray}}
\newcommand{\eeam}{\end{eqnarray}\noindent}

\newcommand{\beqq}{\begin{equation}}
\newcommand{\eeqq}{\end{equation}\noindent}

\newcommand{\bce}{\begin{center}}
\newcommand{\ece}{\end{center}}

\newcommand{\barr}{\begin{array}}
\newcommand{\earr}{\end{array}}

\begin{document}

\title {{\em Preliminary Version} \\ Evidential Value in ANOVA Results in Favor of Fabrication}
\author{
Chris A.J. Klaassen,\\
Korteweg-de Vries Institute for Mathematics \\
University of Amsterdam\\
P.O. Box 94248, 1090 GE Amsterdam, The Netherlands\\
 email: c.a.j.klaassen@uva.nl}

\maketitle

\begin{abstract}
Some scientific publications are under suspicion of fabrication of
data. Since humans are bad random number generators, there might
be some evidential value in favor of fabrication in the
statistical results as presented in such papers. In line with Uri
Simonsohn (2012, 2013) we study the evidential value of the
results of an ANOVA study in favor of the hypothesis of a
dependence structure in the underlying data.
\end{abstract}

\section{Evidential Value in Forensic Statistics}\label{EVFS}

At some crime scene a trace has been found that links a suspect to
the crime. In the court case the prosecutor puts forward the
hypothesis $H_p$ that the suspect is the donor of the trace. The
defendant claims the hypothesis $H_d$ holds, which states that an
unknown person, not the suspect, is the donor of the trace. The
juror (judge, jury) has to decide in favor of $H_p$ or $H_d.$ An
important current scientific approach to such criminal court cases
is via the so-called {\em Bayesian Paradigm of Forensic
Statistics.}

Within this paradigm the juror has to construct a prior opinion
about $H_p$ and $H_d.$ This means that the juror has to decide
beforehand, before seeing the evidence, how likely the hypothesis
of the prosecutor is in comparison to the hypothesis of the
defendant. This prior opinion might be based on e.g. the number of
possible offenders, and it may be formulated in terms of the prior
odds in favor of the hypothesis of the prosecutor, namely
$$ P(H_p)\,/\,P(H_d).$$
The evidence in such a court case consists of the trace found at
the crime scene and characteristics of the suspect. Let us denote
it by $E.$ The forensic expert has to determine now the
probability that a randomly chosen person would leave a trace like
the one found, at the crime scene. This probability is denoted by
$P(E\,|\,H_d).$ Likewise he has to determine $P(E\,|\,H_p),$ the
probability that the suspect would leave a trace like the one
found, at the crime scene. The ratio
$$P(E\,|\,H_p)\,/\,P(E\,|\,H_d)$$ is called the likelihood ratio.
Multiplying the prior odds and the likelihood ratio the juror
obtains the so-called posterior odds in favor of the hypothesis of
the prosecutor $$P(H_p\,|\,E)\,/\,P(H_d\,|\,E),$$ i.e., the odds
in favor of $H_p$ after having seen the evidence. The juror has to
base his decision on these posterior odds. In summary, the
Bayesian Paradigm of Forensic Statistics reads as follows
\begin{equation}\label{BayesParadigm}
 \underbrace{\frac{P(H_p)}{P(H_d)}}_{prior\, odds}\ \
\overbrace{\frac{P(E\,|\,H_p)}{P(E\,|\,H_d)}}^{likelihood\,
ratio}\, =\
\underbrace{\frac{P(H_p\,|\,E)}{P(H_d\,|\,E)}}_{posterior\,
odds}\,.
\end{equation}\\
The validity of equation (\ref{BayesParadigm}) may be checked
straightforwardly by applying the definition of conditional
probability, which is $$P(A\,|\,B) = P(A \cap B)\,/\,P(B),$$ where
$A \cap B$ is the intersection of $A$ and $B.$ Since the
likelihood ratio in (\ref{BayesParadigm}) may be interpreted as
the weight that the evidence should have in the decision of the
juror, it is often called the evidential value in forensic
science.

The evidence $E$ is viewed here as a realization of a random
mechanism, both under $H_d$ and $H_p.$ In case this random
mechanism produces outcomes via probability density functions
$f(E\,|\,H_p)$ and $f(E\,|\,H_d),$ the probabilities in the
likelihood ratio or evidential value are replaced by the
corresponding probability density functions, resulting in
\begin{equation}\label{BayesParadigmdensity}
 \underbrace{\frac{P(H_p)}{P(H_d)}}_{prior\, odds}\ \
\overbrace{\frac{f(E\,|\,H_p)}{f(E\,|\,H_d)}}^{likelihood\,
ratio}\, =\
\underbrace{\frac{P(H_p\,|\,E)}{P(H_d\,|\,E)}}_{posterior\,
odds}\,.
\end{equation}\\

\section{Modelling Fabrication of Data Underlying an ANOVA Study}\label{MFD}

In Analysis of Variance the basic assumption is that all
observations may be viewed as realizations of independent normally
distributed random variables with the same variance $\sigma^2$ and
with means that depend on the values of some categorical
covariates. Let $I$ be the total number of cells that are defined
via these categorical covariates, and let the number of
observations per cell be the same, namely $n.$ The random
variables denoting the observations are then
\begin{equation}\label{model}
X_{ij}= \mu_i +\varepsilon_{ij}\,,\quad i=1,\dots,I,\, j=1,\dots,
n.
\end{equation}
The cell means $\mu_i$ are unknown real numbers, and the
measurement errors $\varepsilon_{ij}$ are independent, normally
distributed random variables with mean 0 and variance $\sigma^2.$

If authors are fiddling around with data and are fabricating and
falsifying data, they tend to underestimate the variation that the
data should show due to the randomness within the model. Within
the framework of the above ANOVA case, we model this by
introducing dependence between the normal random variables
$\varepsilon_{ij},$ which represent the measurement errors.
Actually, we assume that the measurement errors in any cell have
correlation coefficient $\rho$ with respect to the corresponding
measurement errors in the other cells. More precisely formulated,
we assume that the correlations between the random variables
$\varepsilon_{ij}$ no longer all vanish, but satisfy
\begin{equation}\label{dependence}
\rho(\varepsilon_{ij}, \varepsilon_{hj})= \rho,\quad
j=1,\dots,n,\, 0\leq i \neq h \leq I,
\end{equation}
with all other correlations still being equal to 0. In the sequel
we restrict attention to nonnegative values of $\rho$ and we
exclude $\rho=1$ for technical reasons, so $0 \leq \rho < 1.$ We
note that under the standard assumptions of ANOVA $\rho=0$ holds.
Furthermore, we note that within cells observations may be
renumbered in order to get the structure (\ref{dependence}).
Nevertheless, we still assume (\ref{model}) to hold and the
measurement errors to be normally distributed with mean 0 and
variance $\sigma^2.$

A way in which fabrication of measurement errors may take place is
by copying some of them. This might be modelled as follows. Let
$U_j\,,\, j=1, \dots, n,$ and $V_{ij}\,,\, i=1, \dots, I,\ j=1,
\dots, n,$ be independent and identically distributed normal
random variables with mean 0 and variance $\sigma^2.$ Independent
of these, let the random indicators $\Delta_{ij}\,,\, i=1, \dots,
I,\ j=1, \dots, n,$ be independent and identically distributed
Bernoulli random variables with $P(\Delta_{ij}=1)=\sqrt \rho$ and
$P(\Delta_{ij}=0)=1- \sqrt \rho.$ Then
\begin{equation}\label{structure}
\varepsilon_{ij}=\Delta_{ij} U_j + (1-\Delta_{ij}) V_{ij}\,,\quad
i=1, \dots, I,\ j=1, \dots, n,
\end{equation}
satisfy (\ref{dependence}) and (\ref{model}). Note that for $0\leq
i \neq h \leq I$ we have $\varepsilon_{ij}= \varepsilon_{hj}= U_j$
with probability ${\sqrt \rho}^2 = \rho$ then, and the measurement
errors satisfy (\ref{dependence}).

Finally, we note that (\ref{dependence}) is just one possible way
to model dependence, and that the actual way in which fabrication
has been implemented, might lead to quite different dependence
structures. However, this model will come close to some types of
fabrication and falsification.

\section{Evidential Value for Fabrication of Data Underlying an ANOVA Study}\label{EVFD}

Consider a study in a scientific research paper. The data in this
study are analyzed by ANOVA and presented via the sample averages
of the cells and the values of some F-statistics. The underlying
data themselves are not published and are not available. The
conclusion of this study is that the $I$ cells can be grouped into
$K$ groups of $I_k$ cells ($\sum_{k=1}^K I_k=I$), such that
(possibly after renumbering of the cells) group $k$ consists of
cells $i=L_{k-1}+1, \dots, L_k,\, k=1,\dots,K,$ with $0=L_0 < L_1<
\dots < L_K=I,\ L_k -L_{k-1}=I_k,$ and such that for each group
the population cell means are the same, i.e.,
\begin{equation}\label{conclusion}
\mu_i=\nu_k,\quad i=L_{k-1}+1,\dots,L_k,\quad k=1,\dots,K,
\end{equation}
for some values $\nu_k,\, k=1,\dots,K.$

There are two hypotheses to be formulated about the data
underlying the ANOVA study. The hypothesis $H_p$ of fabrication of
the data underlying the results presented in the paper, is
$0<\rho<1.$ The other hypothesis $H_d$ represents the situation
that data have been collected according to (\ref{model}) with
independent $X_{ij},$ i.e., $\rho=0.$ We want to determine the
evidential value of the ANOVA study, i.e., of the sample means of
the cells and the published F-statistics, with respect to these
hypotheses $H_p$ and $H_d.$

To this end we first note that the sample averages in the cells,
\begin{equation}\label{sampleaverage}
X_{i\cdot}=\frac 1n \sum_{j=1}^n X_{ij},\quad i=1,\dots,I,
\end{equation}
have a joint $I$-dimensional multivariate normal distribution.
Actually, the dependence structure (\ref{dependence}) implies
\begin{equation}\label{multinormal}
\begin{pmatrix}
X_{1\,\cdot} \cr \cdot \cr \cdot \cr X_{I\,\cdot}\cr
\end{pmatrix}
\sim \cal N \left(
\begin{pmatrix}
\mu_1 \cr \cdot \cr \cdot \cr \mu_I\cr
\end{pmatrix}
,\, \mbox{$\sigma^2 n^{-1}$}
\begin{pmatrix} 1& \rho & \cdot & \rho \cr \rho & 1 & \cdot & \rho \cr
\cdot & \cdot & \cdot & \cdot \cr \rho & \rho & \cdot & 1 \cr
\end{pmatrix}
\right).
\end{equation}
In stead of assuming (\ref{dependence}), we could have started
right away from (\ref{multinormal}).

Since the inverse of the covariance matrix in (\ref{multinormal})
equals
\begin{equation}\label{inverse}
\mbox{\Large{$\frac n{\sigma^2 (1-\rho)(1+(I-1)\rho)}$}}
\begin{pmatrix} 1+(I-2)\rho& -\rho & \cdot & -\rho \cr -\rho & 1+(I-2)\rho & \cdot & -\rho \cr
\cdot & \cdot & \cdot & \cdot \cr -\rho & -\rho & \cdot &
1+(I-2)\rho \cr
\end{pmatrix}
\end{equation}
and the determinant of $n\sigma^{-2}$ times this covariance matrix
equals
\begin{eqnarray}\label{determinant}
\lefteqn{
 \left|
\begin{array}{cccc}
    1 & \rho & \cdot & \rho \\
    \rho    & 1 & \cdot & \rho \\
    \cdot & \cdot & \cdot & \cdot \\
    \rho & \rho & \cdot & 1
\end{array} \right|
= \left|
\begin{array}{cccc}
    1 & \rho & \cdot & \rho \\
    \rho -1    & 1 -\rho & \cdot & 0 \\
    \cdot & \cdot & \cdot & \cdot \\
    \rho-1 & 0 & \cdot & 1-\rho
\end{array} \right| }\\
&&= \left|
\begin{array}{cccc}
    1+(I-1)\rho & \rho & \cdot & \rho \\
    0    & 1 -\rho & \cdot & 0 \\
    \cdot & \cdot & \cdot & \cdot \\
    0 & 0 & \cdot & 1-\rho
\end{array} \right| =  \left(1+(I-1)\rho \right)
(1-\rho)^{I-1},\nonumber
\end{eqnarray}
(\ref{multinormal}) and (\ref{conclusion}) entail that the joint
density of $X_{1\cdot},\dots,X_{I\cdot}$ equals
\begin{eqnarray}\label{jointdensity1}
\lefteqn{\left(\frac n{2\pi \sigma^2}\right)^{I/2}
\left[(1+(I-1)\rho)(1-\rho)^{I-1}\right]^{-1/2}} \nonumber
\\
 && \exp\left(-\frac n{2\sigma^2} \left[
 \frac 1{1-\rho}\sum_{k=1}^K\sum_{i=L_{k-1}+1}^{L_k}
 \left(X_{i\cdot}-\nu_k\right)^2
 \right. \right.  \\
&&  \qquad \qquad \qquad \left. \left.
  -\, \frac \rho {(1+(I-1)\rho)(1-\rho)} \left(\sum_{k=1}^K
 \sum_{i=L_{k-1}+1}^{L_k}\left(X_{i\cdot}-\nu_k\right)\right)^2
\right]\right). \nonumber
\end{eqnarray}
This density depends on the parameters $\rho, \sigma^2,
\nu_1,\dots, \nu_K.$ If the underlying data would be available
their mean square error
\begin{equation}\label{MSE}
\frac 1{I(n-1)} \sum_{i=1}^I
\sum_{j=1}^n\left(X_{ij}-X_{i\cdot}\right)^2
\end{equation}
would be the proper unbiased estimator of $\sigma^2.$ The
distribution of this estimator depends on $\rho,$ but its mean
does not. Furthermore, standard ANOVA theory shows that this
estimator is independent of the exponent in (\ref{jointdensity1}).
Since the underlying data are not available, the value of the
parameter $\sigma^2$ should be retrieved from the values of the
F-statistics given. For a method to do this that does not depend
on $\rho,$ see the next section. Let us call the resulting
estimate ${\hat \sigma}_n^2,$ and let us denote the density from
(\ref{jointdensity1}) with $\sigma$ replaced by ${\hat \sigma}_n$
by $f_n(X_{1\cdot},\dots, X_{I\cdot}; \nu_1,\dots,\nu_K,\rho).$

The hypothesis $H_p$ of fabrication of the data corresponds to the
parameter values $0<\rho<1$ and $\nu_1,\dots,\nu_K$ arbitrary, and
the hypothesis $H_d$ of proper data corresponds to the parameter
values $\rho = 0$ and $\nu_1,\dots,\nu_K$ arbitrary. The
evidential value
$$\frac{f(E\,|\,H_p)}{f(E\,|\,H_d)}$$ from
(\ref{BayesParadigmdensity}) in favor of $H_p$ versus $H_d$
becomes in this case (cf. Zhang (2009), Bickel (2012))
\begin{equation}\label{evidentialvalue1}
\mathbb V = \frac{\sup_{0<\rho<1,\, \nu_1,\dots,\nu_K \in \mathbb
R}f_n(X_{1\cdot},\dots, X_{I\cdot}; \nu_1,\dots,\nu_K,\rho)}
{\sup_{\nu_1,\dots,\nu_K \in \mathbb R}f_n(X_{1\cdot},\dots,
X_{I\cdot}; \nu_1,\dots,\nu_K, 0)}.
\end{equation}
Straightforward computation shows that for any $\rho$
\begin{equation}\label{nus}
\sup_{\nu_1,\dots,\nu_K \in \mathbb R}f_n(X_{1\cdot},\dots,
X_{I\cdot}; \nu_1,\dots,\nu_K,\rho)
\end{equation}
is attained at
$$\nu_k= {\bar X}_k =\frac 1{L_k -L_{k-1}}
\sum_{i=L_{k-1}+1}^{L_k} X_{i\cdot}\,,\quad k=1,\dots,K.$$ This
implies that the evidential value from (\ref{evidentialvalue1})
reduces to
\begin{equation}\label{evidentialvalue2}
\mathbb V = \sup_{0<\rho<1} \chi_n(\rho)
\end{equation}
with
\begin{eqnarray}\label{jointdensity2}
\lefteqn{\chi_n(\rho) =
\left[(1+(I-1)\rho)(1-\rho)^{I-1}\right]^{-1/2}}
\\
 &&  \qquad \quad \exp\left(-\frac {n \rho}{2{\hat \sigma}_n^2(1-\rho)}\
 \sum_{k=1}^K \sum_{i=L_{k-1}+1}^{L_k}\left(X_{i\cdot}-{\bar X}_k\right)^2
 \right). \nonumber
\end{eqnarray}
We need the additional notation
\begin{eqnarray}\label{notation}
S_n & = & \frac n{I{\hat \sigma}_n^2}\ \sum_{k=1}^K
\sum_{i=L_{k-1}+1}^{L_k
}\left(X_{i\cdot}-{\bar X}_k\right)^2,
\\
{\hat \rho}_n & = & \frac 12 (1-S_n)\left[1+\sqrt{1-\frac
{4S_n}{(I-1)(1-S_n)^2}}\right].
\end{eqnarray}
In Proposition \ref{computationV} of the Appendix the following is
shown.
\begin{itemize}
\item
If $$S_n \geq \frac {\sqrt I -1}{\sqrt I +1}$$ holds, then the
evidential value from (\ref{evidentialvalue1}) and
(\ref{evidentialvalue2}) reduces to $\mathbb V = 1.$
\item
If $$S_n < \frac {\sqrt I -1}{\sqrt I +1}$$ holds, then ${\hat
\rho}_n$ is well-defined and the evidential value from
(\ref{evidentialvalue1}) and (\ref{evidentialvalue2}) reduces to
\begin{equation}\label{evidentialvalue5}
\mathbb V = \max\left\{\chi_n({\hat \rho}_n), 1\right\}.
\end{equation}
\end{itemize}

\section{Estimating $\sigma^2$ from F-Statistics}\label{EFS}

Table 1 in Stapel, Koomen and Van der Pligt (1996) presents the
sample means in a three-way layout ANOVA study with a $3 \times 2
\times 2$ design.

\medskip
\begin{tabular}{|l|l|l|l|}\hline
 Prime type              & Positive & Negative & Irrelevant \\\hline
 Impersonal / Memory     & 2.3      & 3.5      & 2.9        \\
 Impersonal / Impression & 3.4      & 2.5      & 2.9        \\
 Personal / Memory       & 3.3      & 2.3      & 2.9        \\
 Personal / Impression   & 3.5      & 2.5      & 3.0        \\\hline
\end{tabular}
\medskip

The estimate of the error variance $\sigma^2$ is not given.
It should be possible to retrieve this estimate from the value of
any F-statistic. On page 441 of ibid. the value of the F-statistic
for testing three-way interactions is given, namely,
$F(2,326)=3.21.$ We assume that the 338 observations are
approximately uniformly distributed over the 12 cells. This yields
an average of 28.17 observations per cell. Applying e.g. Table
4.5.2 (Analysis of Variance of the Three-Way Layout with $M$
Observations per Cell) of Scheff\'e (1959) we obtain by some
computation that the mean square error for interaction equals
7.769. Dividing this by the value 3.21 of the F-statistic we get
2.420 as the mean square for error, i.e., the estimate for
$\sigma^2.$ However, this is {\em not} the value that we would
have gotten, would we have used the underlying observations, since
the cell means, which are used in the above computation, are given
in very low precision.

In an ANOVA of the upper half of Table 1 in ibid. the two way
interaction terms are tested by an F-statistic with value
$F(2,164)=14.28.$ By Table 4.3.1 of Scheff\'e (1959) a similar
computation as above yields 1.095 as the value of the mean square
for error, based on 170 observations.

Note that a value like 2.3 for a cell mean implies that the actual
value of the cell mean lies in the interval $[2.25, 2.35).$ Using
this rounding off property we may conclude that the first three
F-values given on page 442 of ibid., which have 1 and 164 degrees
of freedom, imply that the value of the mean square for error,
based on 170 observations, lies in the interval $[0.918, 1.218].$
Note that 1.095 belongs to this interval. Averaging the values of
the mean square for error that we get from the last four F-values,
we obtain 1.047 as our estimate.

The F-values presented on page 442 of ibid. that are based on the
second half of Table 1 of ibid., namely $F(2,162)=11.49$ and $F(1,
162)=23.00,$ yield 1.223 and 1.217 as value of the mean square for
error, based on the remaining 168 observations. Averaging yields
1.220. Pooling 1.047 and 1.220 we obtain $${\hat \sigma}_n^2 =
1.134 $$ as our final estimate for $\sigma^2.$ Note that this
deviates considerably from the value 2.42, which has been obtained
from the F-value 3.21 for three-way interaction. Let us presume
here that this is a misprint and that this F-value should have
been something like 6.9.

In order to take care of the rounding off of the values of the
cell means, we have adapted Table 1 in a direction that increases
the double sum in (\ref{jointdensity2}) as much as possible and
that should decrease the evidential value. The resulting table is
given below.

\medskip
\begin{tabular}{|l|l|l|l|}\hline
 Prime type              & Positive & Negative & Irrelevant \\\hline
 Impersonal / Memory     & 2.25     & 3.55     & 2.85       \\
 Impersonal / Impression & 3.35     & 2.55     & 2.85       \\
 Personal / Memory       & 3.25     & 2.25     & 2.85       \\
 Personal / Impression   & 3.55     & 2.55     & 3.05       \\\hline
\end{tabular}
\medskip

Analyzing the same F-statistics as above and performing the same
computations we see that the F-statistics for interactions yield
exactly the same values for the mean square for error. Only the
three F-statistics of the type $F(1,164)$ yield different values.
Averaging the four values for the mean square for error that we
get out of the four F-values related to the upper half of the
table, we arrive at 1.117. The F-statistics for the second half of
the table yield the same estimate 1.220. Pooling 1.117 and 1.220
we obtain $${\hat \sigma}_n^2 = 1.168 $$ as our final estimate for
$\sigma^2$ based on our version of Table 1 of ibid.

\section{Computing Evidential Value}\label{CEV}

Let us group the cells of the tables in the preceding section into
three groups, namely the groups corresponding to the covariate
Prime type with the first two cells in the row Impersonal / Memory
interchanged; $I=12, K=3, I_1=I_2=I_3=4.$ According to the social
psychology theory as put forward in Stapel, Koomen and Van der
Pligt (1996), the participants within these groups should have
similar mean scores. By (\ref{evidentialvalue2}) through
(\ref{evidentialvalue5}) we may compute the evidential value
$\mathbb V$ in favor of the hypothesis $H_p$ that these data have
been fabricated in some way resulting in (\ref{multinormal}) with
$0<\rho<1.$ For the first table from the preceding section, i.e.,
Table 1 from ibid., this yields
$$\mathbb V = 56.88$$
and for the second, adapted table from the preceding section this
yields $$\mathbb V = 1.92.$$

\section{Interpreting Evidential Value}\label{IEV}

With the evidential value $\mathbb V$ defined as in
(\ref{evidentialvalue2}) through (\ref{evidentialvalue5}) the
Bayesian paradigm for criminal court cases
(\ref{BayesParadigmdensity}) becomes
\begin{equation}\label{BayesParadigmV}
 \underbrace{\frac{P(H_p)}{P(H_d)}}_{prior\, odds}\ \
\overbrace{\mathbb V}^{evidential\, value}\, =\
\underbrace{\frac{P(H_p\,|\,E)}{P(H_d\,|\,E)}}_{posterior\,
odds}\,.
\end{equation}
An important principle in criminal court cases is `in dubio pro
reo', which means that in case of doubt the accused is favored. In
science one might argue that the leading principle should be `in
dubio pro scientia', which should mean that in case of doubt a
publication should be withdrawn. Within the framework of this
paper this would imply that if the posterior odds in favor of
hypothesis $H_p$ of fabrication equal at least 1, then the
conclusion should be that $H_p$ is true. So an ANOVA study for
which
\begin{equation}\label{BayesParadigmV2}
 \underbrace{\frac{P(H_p)}{P(H_d)}}_{prior\, odds}\ \
\overbrace{\mathbb V}^{evidential\, value}\, =\
\underbrace{\frac{P(H_p\,|\,E)}{P(H_d\,|\,E)}}_{posterior\, odds}
> 1
\end{equation}
holds, should be rejected and disqualified scientifically.

We conclude with some notes.
\begin{itemize}
\item
ANOVA studies are based on the assumption of normality. Often this
assumption is not satisfied, but the technique is still applied.
This is the case in Stapel et al. (1996), since in Table 1 of
ibid. the measurements are averages of two 7 point Likert scales,
which hardly behave like normal random variables. However, in view
of the central limit theorem cell means like in our basic model
(\ref{multinormal}) behave approximately like (jointly
multivariate) normal random variables.
\item
Note that (\ref{evidentialvalue5}) implies
$$\mathbb V \geq 1.$$
Consequently, within this framework there does not exist
exculpatory evidence. This is reasonable since bad science cannot
be compensated by very good science. It should be very good
anyway.
\item
When a paper contains more than one study based on independent
data, then the evidential values of both studies can and may be
combined into an overall evidential value by multiplication in
order to determine the validity of the whole paper; see the
preceding item.
\item
One may wonder if the way in which the mean square error
(\ref{MSE}) is retrieved from the values of F-statistics,
interferes with the randomness in (\ref{evidentialvalue1}). As
mentioned in Section \ref{EVFD} standard ANOVA theory shows that
this estimator is independent of the exponent in
(\ref{jointdensity1}) and hence (\ref{evidentialvalue1}), provided
the underlying data have a normal distribution; see also item 1.
\end{itemize}

\section{Evidential Value for Fabrication of Data Underlying an ANOVA Study Based on an Alternative Dependence Structure}\label{EVFDA}

In this Section we present an analysis as in Sections \ref{MFD}
and \ref{EVFD}, but under a different dependence structure. Given
the group structure of the cells as presented in the first
paragraph of Section \ref{EVFD} we assume the existence of
$\rho_1,\dots,\rho_K \in [0,1]$ such that
\begin{equation}\label{dependencea} \rho(\varepsilon_{ij},
\varepsilon_{hj})= \rho_k,\quad j=1,\dots,n,\, L_{k-1}+1\leq i
\neq h \leq L_k,\quad k=1,\dots,K,
\end{equation}
hold with all other correlations being equal to 0. This implies
independence between different groups of cells. We note that
(\ref{dependencea}) is just another possible way to model
dependence, and we note again that the actual way in which
fabrication has been implemented, might lead to quite different
dependence structures.

We reconsider the ANOVA study presented via the sample averages of
the cells and the values of some F-statistics. Again the
underlying data themselves are not published and are not
available, and the conclusion of this study is given by
(\ref{conclusion}). There are two hypotheses to be formulated
about the data underlying the ANOVA study. The hypothesis $H_p$ of
fabrication of the data underlying the results, is that at least
one of the $\rho_k$s is positive. The other hypothesis $H_d$
represents the situation that data have been collected according
to (\ref{model}) with independent $X_{ij},$ i.e.,
$\rho_1=\dots=\rho_K=0.$ We want to determine the evidential value
of the ANOVA study, i.e., of the sample means of the cells and the
published F-statistics, with respect to these hypotheses $H_p$ and
$H_d.$ Here the evidential value is defined analogously to
(\ref{evidentialvalue1}) with the supremum taken over $0<\rho_k
<1,\, k=1,\dots,K.$

The sample averages in the cells, $X_{i\cdot}$ from
(\ref{sampleaverage}), have a joint $I$-dimensional multivariate
normal distribution with
\begin{equation}\label{multinormala}
\begin{pmatrix}
X_{L_{k-1}+1\,\cdot} \cr \cdot \cr \cdot \cr X_{L_k\,\cdot}\cr
\end{pmatrix}
\sim \cal N \left(
\begin{pmatrix}
\nu_k \cr \cdot \cr \cdot \cr \nu_k\cr
\end{pmatrix}
,\, \mbox{$\sigma^2 n^{-1}$}
\begin{pmatrix} 1& \rho_k & \cdot & \rho_k \cr \rho_k & 1 & \cdot & \rho_k \cr
\cdot & \cdot & \cdot & \cdot \cr \rho_k & \rho_k & \cdot & 1 \cr
\end{pmatrix}
\right)
\end{equation}
for each $k=1, \dots, K$ and with independence between groups with
different indices $k.$ This entails that the joint density of
$X_{1\cdot},\dots,X_{I\cdot}$ equals
\begin{eqnarray}\label{jointdensity1a}
\lefteqn{\left(\frac n{2\pi \sigma^2}\right)^{I/2} \prod_{k=1}^K
\left[(1+(I_k-1)\rho_k)(1-\rho_k)^{I_k-1}\right]^{-1/2}} \nonumber
\\
 && \exp\left(-\frac n{2\sigma^2} \sum_{k=1}^K\left[
 \frac 1{1-\rho_k}\sum_{i=L_{k-1}+1}^{L_k}
 \left(X_{i\cdot}-\nu_k\right)^2
 \right. \right.  \\
&&  \qquad \qquad \qquad \left. \left.
  -\, \frac {\rho_k} {(1+(I_k-1)\rho_k)(1-\rho_k)} \left(
 \sum_{i=L_{k-1}+1}^{L_k}\left(X_{i\cdot}-\nu_k\right)\right)^2
\right]\right). \nonumber
\end{eqnarray}
This density depends on the parameters $\rho_1,\dots, \rho_K,
\sigma^2, \nu_1,\dots, \nu_K.$ Again, we write ${\hat
\sigma}_n^2,$ for the estimate of $\sigma^2.$

With the notation
\begin{eqnarray}\label{notationa}
\lefteqn{\chi_{n,k}(\rho) =
\left[(1+(I_k-1)\rho)(1-\rho)^{I_k-1}\right]^{-1/2}} \nonumber
\\
 &&  \qquad \quad \exp\left(-\frac {n \rho}{2{\hat \sigma}_n^2(1-\rho)}\
 \sum_{i=L_{k-1}+1}^{L_k}\left(X_{i\cdot}-{\bar X}_k\right)^2
 \right), \nonumber \\
S_{n,k} & = & \frac n{I_k{\hat \sigma}_n^2}\
\sum_{i=L_{k-1}+1}^{L_k }\left(X_{i\cdot}-{\bar X}_k\right)^2,
\\
{\hat \rho}_{n,k} & = & \frac 12 (1-S_{n,k})\left[1+\sqrt{1-\frac
{4S_{n,k}}{(I_k-1)(1-S_{n,k})^2}}\right]\,{\bf 1}_{[S_{n,k} <
(\sqrt{I_k}-1)/(\sqrt{I_k}+1)]}  \nonumber
\end{eqnarray}
Proposition \ref{computationV} of the Appendix shows
\begin{equation}
\mathbb V = \prod_{k=1}^K \chi_{n,k}\left({\hat
\rho}_{n,k}\right).
\end{equation}

Computing this evidential value for Table 1 in Stapel, Koomen and
Van der Pligt (1996), i.e., for the first table of Section
\ref{EFS}, we obtain $$\mathbb V = 14.49.$$ The adapted table,
namely the second table of Section \ref{EFS}, yields
$$\mathbb V = 1.28.$$

Here and in (\ref{evidentialvalue1}) we have defined the
evidential value in the presence of the nuisance parameters
$\nu_1,\dots, \nu_K$ by replacing these parameters by their
maximum likelihood estimators. An alternative approach is to
compute the evidential value keeping these parameters fixed, and
to subsequently minimize the resulting evidential value over these
nuisance parameters; in formula
\begin{equation}\label{evidentialvalue6}
\tilde {\mathbb V} = \inf_{\nu_1,\dots,\nu_K \in \mathbb R}
\frac{\sup_{0<\rho_k<1, k=1,\dots,K}f_n(X_{1\cdot},\dots,
X_{I\cdot}; \nu_1,\dots,\nu_K,\rho_1,\dots,\rho_K)}
{f_n(X_{1\cdot},\dots, X_{I\cdot}; \nu_1,\dots,\nu_K, 0,\dots,0)},
\end{equation}
where $f_n(X_{1\cdot},\dots, X_{I\cdot};
\nu_1,\dots,\nu_K,\rho_1,\dots,\rho_K)$ is the density as given in
(\ref{jointdensity1a}) with $\sigma$ replaced by ${\hat
\sigma}_n.$ In fact, both definitions of evidential value yield
the same value in the situation of this Section \ref{EVFDA}, as is
shown in Theorem \ref{V=V}.

\appendix

\section{Appendix: Analysis $\chi$ Function}
Here we present a proof of the main result of Section \ref{EVFD}.
\begin{prop}\label{computationV}
In the notation (\ref{jointdensity2}) and (\ref{notation}) and for
$I\geq 2$
\begin{equation}\label{result}
\sup_{0<\rho <1} \chi_n(\rho)= {\bf 1}_{[S_n \geq (\sqrt I
-1)/(\sqrt I +1)]} + \max\left\{\chi_n({\hat \rho}_n),
1\right\}\,{\bf 1}_{[S_n < (\sqrt I -1)/(\sqrt I +1)]}
\end{equation}
holds.
\end{prop}

\noindent{\bf Proof}\\
Write $\psi_n(\rho)=\log\left(\chi_n(\rho)\right),\, 0\leq \rho
<1,$ and $\psi_n'(\rho)$ for its derivative. Some computation
shows that
\begin{equation}\label{result1}
\psi_n(0)=0,\quad \psi_n'(0)=-\frac 12 I S_n,
\end{equation}
hold and that $\psi_n'(\rho)$ is nonnegative on the interval
$[0,1)$ if and only if both
$$S_n \leq (\sqrt I -1)/(\sqrt I +1)$$
and
\begin{eqnarray}\label{result2}
\lefteqn{\frac 12 (1-S_n)\left[1-\sqrt{1-\frac
{4S_n}{(I-1)(1-S_n)^2}}\, \right]}\\
&& \leq \rho \leq \frac 12 (1-S_n)\left[1+\sqrt{1-\frac
{4S_n}{(I-1)(1-S_n)^2}}\, \right] ={\hat \rho}_n \nonumber
\end{eqnarray}
hold. Consequently, $\psi_n(\rho)$ and $\chi_n(\rho)$ have (local)
maxima at $\rho=0$ and $\rho={\hat \rho}_n$ on $[0,1).$ This
implies (\ref{result}).
\hfill$\Box$\\

\section{Appendix: Alternative Definition of Evidential Value}

The alternative definition (\ref{evidentialvalue6}) of evidential
value yields the same value as (\ref{evidentialvalue1}) for the
alternative dependence model as given in Section \ref{EVFDA}.

\begin{thm}\label{V=V}
In the situation of Section \ref{EVFDA} the evidential values as
defined by (\ref{evidentialvalue1}) and (\ref{evidentialvalue6})
satisfy
\begin{equation}\label{V=Vtilde}
\tilde{\mathbb V}=\mathbb V.
\end{equation}
\end{thm}

\noindent{\bf Proof}\\
First we note
\begin{eqnarray}\label{evidentialvalue7}
\tilde {\mathbb V} & = & \inf_{\nu_1,\dots,\nu_K \in \mathbb R}
\frac{\sup_{0<\rho<1}f_n(X_{1\cdot},\dots, X_{I\cdot};
\nu_1,\dots,\nu_K,\rho)} {f_n(X_{1\cdot},\dots, X_{I\cdot};
\nu_1,\dots,\nu_K, 0)} \nonumber \\
& \leq & \inf_{\nu_1,\dots,\nu_K \in \mathbb R}
\frac{\sup_{0<\rho<1,\, \nu_1^*,\dots,\nu_K^* \in \mathbb
R}f_n(X_{1\cdot},\dots, X_{I\cdot}; \nu_1^*,\dots,\nu_K^*,\rho)}
{f_n(X_{1\cdot},\dots,
X_{I\cdot}; \nu_1,\dots,\nu_K, 0)} \nonumber \\
& = & \frac{\sup_{0<\rho<1,\, \nu_1,\dots,\nu_K \in \mathbb
R}f_n(X_{1\cdot},\dots, X_{I\cdot}; \nu_1,\dots,\nu_K,\rho)}
{\sup_{\nu_1,\dots,\nu_K \in \mathbb R}f_n(X_{1\cdot},\dots,
X_{I\cdot}; \nu_1,\dots,\nu_K, 0)}=\mathbb V.
\end{eqnarray}
Subsequently, we note that by the product structure of
(\ref{jointdensity1a}) it suffices to consider the case $K=1$ in
proving $\tilde {\mathbb V} \geq \mathbb V.$ Furthermore, by
(\ref{jointdensity1a}) with $K=1$ we have
\begin{eqnarray}\label{evidentialvalue8}
\lefteqn{\frac{\sup_{0<\rho<1}f_n(X_{1\cdot},\dots, X_{I\cdot};
\nu,\rho)} {f_n(X_{1\cdot},\dots, X_{I\cdot};
\nu, 0)} \nonumber} \\
& = & \sup_{0<\rho<1} \left[(1+(I-1)\rho)(1-\rho)^{I-1}\right]^{-1/2} \nonumber \\
&& \qquad \exp\left(-\frac {n\rho}{2{\hat \sigma}_n^2(1-\rho)}
\left[\sum_{i=1}^I \left(X_{i\cdot}-\nu\right)^2 \right. \right.  \\
&&  \qquad \qquad \qquad \left. \left.-\, \frac 1 {1+(I-1)\rho}
\left( \sum_{i=1}^I\left(X_{i\cdot}-\nu\right)\right)^2
\right]\right). \nonumber
\end{eqnarray}
With the notation $\bar X=I^{-1}\sum_{i=1}^I X_{i\,\cdot}$ we
obtain
\begin{eqnarray}\label{inequality}
\lefteqn{\sum_{i=1}^I \left(X_{i\cdot}-\nu\right)^2 -\, \frac 1
{1+(I-1)\rho} \left(
\sum_{i=1}^I\left(X_{i\cdot}-\nu\right)\right)^2} \nonumber \\
&&=\sum_{i=1}^I \left(X_{i\cdot}-\bar X\right)^2 +\left(I- \frac
{I^2}{1+(I-1)\rho}\right)(\bar X -\nu)^2 \\
&& \leq \sum_{i=1}^I \left(X_{i\cdot}-\bar X\right)^2 \nonumber
\end{eqnarray}
in view of $\rho <1.$ Together with (\ref{evidentialvalue8}) this
inequality yields
\begin{eqnarray}\label{evidentialvalue9}
\lefteqn{\tilde {\mathbb V} \geq \inf_\nu
\sup_{0<\rho<1} \left[(1+(I-1)\rho)(1-\rho)^{I-1}\right]^{-1/2}} \nonumber \\
&& \qquad \exp\left(-\frac {n\rho}{2{\hat \sigma}_n^2(1-\rho)}
\sum_{i=1}^I \left(X_{i\cdot}-\bar X\right)^2 \right).
\end{eqnarray}
Since the infimum over $\nu$ may be removed from
(\ref{evidentialvalue9}), equations (\ref{evidentialvalue2}) and
(\ref{jointdensity2}) with $K=1$ imply $\tilde {\mathbb V} \geq
\mathbb V,$ which completes the proof.
\hfill$\Box$\\

\section{Appendix: $t$-Statistic under Dependence}
If one would be interested in the distribution of the exponent in
(\ref{jointdensity2}) or of $S_n$ from (\ref{notation}), the
following lemma would come in handy.

\begin{lem}\label{chisquared}
Let the correlated standard normal random variables
$Z_1,\dots,Z_d$ have a joint multivariate normal distribution,
namely
\begin{equation}\label{multinormal2}
Z=
\begin{pmatrix}
Z_1 \cr \cdot \cr \cdot \cr Z_d \cr
\end{pmatrix}
\sim \cal N \left( \mbox{$0$} ,
\begin{pmatrix} 1& \rho & \cdot & \rho \cr \rho & 1 & \cdot & \rho \cr
\cdot & \cdot & \cdot & \cdot \cr \rho & \rho & \cdot & 1 \cr
\end{pmatrix}
\right)
\end{equation}
with $0 \leq \rho <1.$ Let
\begin{equation}\label{samplestat} {\bar Z}_d =
\frac 1d \sum_{i=1}^d Z_i\,,\quad S_d^2 = \frac 1{d-1}
\sum_{i=1}^d \left(Z_i- {\bar Z}_d \right)^2
\end{equation}
be their sample mean and sample variance, respectively.

Then ${\bar Z}_d$ and $S_d^2$ are independent, ${\bar Z}_d$ has a
normal distribution with mean 0 and variance $(1+(d-1)\rho)/d,$
and $(d-1)S_d^2/(1-\rho)$ has a chi squared distribution with
$d-1$ degrees of freedom.
\end{lem}

\noindent{\bf Proof}\\
The following classical trick for the case $\rho=0$ also works for
positive $\rho.$ Let $A^T$ be an orthogonal (orthonormal) matrix,
the first row of which is the row vector $(d^{-1/2}, \dots,
d^{-1/2}).$ Define the column $d$-vector $Y$ by $Y=A^T Z,$ and
note
\begin{eqnarray}\label{relYZ}
Y_1&=&d^{1/2}{\bar Z}_d\,, \nonumber \\
(d-1)S_d^2&=&Z^T Z-d {\bar Z}_d^2= Y^T A^T AY- Y_1^2 \nonumber \\
&=& Y^T Y-Y_1^2= \sum_{i=2}^d Y_i^2\,, \\
EY &=& 0, \nonumber
\end{eqnarray}
and
\begin{eqnarray}\label{covY}
 A^T
\begin{pmatrix} 1& \rho & \cdot & \rho \cr \rho & 1 & \cdot & \rho \cr
\cdot & \cdot & \cdot & \cdot \cr \rho & \rho & \cdot & 1 \cr
\end{pmatrix}
A& =& A^T \left((1-\rho)
\begin{pmatrix} 1& 0 & \cdot & 0 \cr 0 & 1 & \cdot & 0 \cr
\cdot & \cdot & \cdot & \cdot \cr 0 & 0 & \cdot & 1 \cr
\end{pmatrix}
+
\begin{pmatrix} \rho& \rho & \cdot & \rho \cr \rho & \rho & \cdot & \rho \cr
\cdot & \cdot & \cdot & \cdot \cr \rho & \rho & \cdot & \rho \cr
\end{pmatrix}
\right) A \nonumber \\
&=&(1-\rho)
\begin{pmatrix} 1& 0 & \cdot & 0 \cr 0 & 1 & \cdot & 0 \cr
\cdot & \cdot & \cdot & \cdot \cr 0 & 0 & \cdot & 1 \cr
\end{pmatrix}
+
\begin{pmatrix} {\sqrt d}\rho& {\sqrt d}\rho & \cdot & {\sqrt d}\rho \cr
0 & 0 & \cdot & 0 \cr \cdot & \cdot & \cdot & \cdot \cr 0 & 0 &
\cdot & 0 \cr
\end{pmatrix}
A \nonumber \\
&=&(1-\rho)
\begin{pmatrix} 1& 0 & \cdot & 0 \cr 0 & 1 & \cdot & 0 \cr
\cdot & \cdot & \cdot & \cdot \cr 0 & 0 & \cdot & 1 \cr
\end{pmatrix}
+
\begin{pmatrix} d\rho& 0 & \cdot & 0 \cr
0 & 0 & \cdot & 0 \cr \cdot & \cdot & \cdot & \cdot \cr 0 & 0 &
\cdot & 0 \cr
\end{pmatrix}
\nonumber \\
&=&
\begin{pmatrix} 1+(d-1)\rho& 0 & \cdot & 0 \cr
0 & 1-\rho & \cdot & 0 \cr \cdot & \cdot & \cdot & \cdot \cr 0 & 0
& \cdot & 1-\rho \cr
\end{pmatrix} ,
\end{eqnarray}
where the matrix equalities hold because $A^T A$ equals the
identity matrix and because all row vectors of $A^T$ are
orthogonal to its first row vector $(d^{-1/2}, \dots, d^{-1/2}),$
and hence to all multiples of $(1,\dots,1).$ Since (\ref{covY}) is
the covariance matrix of the multivariate normally distributed
vector $Y,$ it follows that $Y_1, \dots, Y_d$ are independent, and
consequently, that ${\bar Z}_d$ and $S_d^2$ are. Finally,
(\ref{relYZ}) and (\ref{covY}) imply that $Y_2, \dots, Y_d$ are
independent identically distributed with a normal distribution
with mean 0 and variance $1-\rho,$ which yields that
$(1-\rho)^{-1} \sum_{i=2}^d Y_i^2$ has a chi squared distribution
with $d-1$ degrees of freedom.
\hfill$\Box$\\

We note that as a consequence the statistic
\begin{equation}\label{tstat}
\sqrt{\frac{d(1-\rho)}{1+(d-1)\rho}}\ \frac {{\bar Z}_d}{S_d}
\end{equation}
has a $t$-distribution with $d-1$ degrees of freedom.\\

\noindent{\bf Acknowledgement} \\
\noindent We would like to thank Marjan Sjerps, Bert van Es, and
Uri Simonsohn for useful discussions and Uri Simonsohn for sharing
his ideas as presented in Simonsohn (2012, 2013).

\end{document}